%% file: MN_15_4180_L_clean.tex
\documentclass[fleqn,usenatbib]{mnras}

\usepackage{pict2e}
\usepackage[T1]{fontenc}
\usepackage{ae,aecompl}

\usepackage{graphicx}	% Including figure files
\usepackage{amsmath}	% Advanced maths commands
\usepackage{amssymb}	% Extra maths symbols
\usepackage{xcolor}

\input{newcommands.tex}

\graphicspath{{Images/}{Images/Periodogram/}}

\title[QPO signatures in the optical polarization of \pks]{Evidence for quasi-periodic oscillations in the optical polarization of the blazar \pks}

\author[N.~W. Pekeur et al.]{N.\ W.\ Pekeur$^{1}$\thanks{Contact e-mail: \href{mailto:nikki.pekeur@ast.uct.ac.za}{nikki.pekeur@ast.uct.ac.za}}, A.\ R.\ Taylor$^{1,2}$, S.\ B. Potter$^3$ and R.\ C.\ Kraan-Korteweg$^1$
\\
$^{1}$Astrophysics, Cosmology and Gravity Centre, Department of Astronomy, University of Cape Town, Private Bag X3, Rondebosch 7701, SA
\\
$^{2}$Department of Physics and Astronomy, University of the Western Cape, Modderdam Road, Private Bag X17, Belville 7530, SA
\\
$^{3}$South African Astronomical Observatory, PO Box 9,
  Observatory 7935, Cape Town, SA}

\date{Last updated 2015 May 22; in original form 2013 September 5}

\pubyear{2015}

\begin{document}
\label{firstpage}
\pagerange{\pageref{firstpage}--\pageref{lastpage}}
\maketitle

\begin{abstract}
Evidence for the presence of quasi-periodic oscillations (QPOs) in the optical polarization of the blazar \pks, during a period of enhanced gamma-ray brightness, is presented. The periodogram of the polarized flux revealed the existence of a prominent peak at $T\sim 13$ min, detected at $>99.7$\% significance, and $T\sim 30$ min, which was nominally significant at $>99$\%. This is the first evidence of QPOs in the polarization of an active galactic nucleus, potentially opening up a new avenue of studying this phenomenon.
\end{abstract}


%%%%%%%%%%%%%%%%%%%%%%%%%%%%%%%%%%%%%%%%%%%%%%%%%%

%%%%%%%%%%%%%%%%% BODY OF PAPER %%%%%%%%%%%%%%%%%%

\section{Introduction}
\pks{} is a blazar, a subclass of radio-loud active galactic nucleus (AGN) for which the relativistic jet is oriented close to the line-of-sight of the observer \citep{Urry2009}. Active galactic nuclei are powered by accretion of matter onto a $10^6-10^9$\msun{} supermassive black hole (SMBH). The observed emission is dominated by the jet and is characterized by intense and rapid brightness fluctuations across the electromagnetic spectrum, and the presence of polarization at optical to radio wavelengths. \pks{} exhibits significant variability on timescales ranging from minutes \citep{HESS2007} to years \citep{Fan2000, Kastendieck2011}. Short timescale variations, from minutes to hours,  probe emission regions with sizes comparable to the gravitational radius of the central SMBH, while variations on timescales of months to years probe the jet structure. The timing signatures indicate that the emission is governed by correlated noise processes \citep{Vaughan2003, Edelson1999, Vaughan2005, Chatterjee2008, Chatterjee2012, Kastendieck2011}, possibly originating from instabilities in the accretion rate and jet. 

Quasi-periodic variations have been reported for a small number of AGN. The clearest detection is a $\sim 1$ hour QPO in the \xray{} light curve of the radio-quiet Seyfert galaxy RE J1034+396 \citep{GMWD2008}. \pks{} is one of only a few blazars for which convincing evidence of quasi-periodic brightness variations have been documented. Long time-scale periodicities were discovered in the optical light curve of the source. Amplitude modulations with periods of 4 and 7 years \citep{Fan2000}, and 315 days \citep{Zhang2014, Sandrinelli2014} were identified, while a short time-scale oscillation of \til 4.6 hours was observed at \xray{}energies \citep{Lachowicz2009}.

Evidence of quasi-periodic oscillations in the optical polarization of \pks{} is presented here. The polarization of the source was monitored from 25 -- 27 July, 2009 with the HIgh Speed Photo-POlarimeter (HIPPO) of the Southern African Astronomical Society (SAAO). Gamma-ray observations were obtained independently with the High Energy Stereoscopic System (\hess) over the same period, with indications of two \gray{} flares which bracket the occurrence of the optical polarization fluctuations. The acquisition of the optical polarization measurements and a description of the polarization and \gray{} observations is given in section~\ref{sect:obs}. Section~\ref{sect:analysis} presents an analysis of the observations. Discussion of the results follows in section~\ref{sect:discussion}, while the conclusions are presented in section~\ref{sect:conclusions}. 

%cites relevant earlier studies in the field by \citet{Others2013}, and describes the problem the authors aim to solve \citep[e.g.][]{Author2012}.

\section{Observations}\label{sect:obs}
The optical polarization of \pks{} was observed with the HIPPO \citep{HIPPO}, a rapidly rotating dual-channel photo-polarimeter operated by the SAAO. The instrument is mounted on the 1.9 metre Radcliffe telescope, which is located in Sutherland (South Africa). A comprehensive description of the polarimeter, data acquisition and reduction is presented in \citealt{HIPPO}. The most important instrument characteristics are described below. 

The polarization of the incident light is measured by passing the incoming beam through a $\frac{1}{4}$ waveplate and a $\frac{1}{2}$ waveplate, which detects circular and linear polarization, respectively. The waveplates contra-rotate at 10 Hz, producing a modulation of the intensity dependent on the polarization of the beam.  Simultaneous measurement of all Stokes parameters is obtained by rapidly sampling the polarization ellipse of the incoming beam as a function of angles during the rotation of the wave plates. A Thompson beam-splitter separates the beam into the orthogonal components of the electric field vector components, thereby allowing two independent measurements of the polarization. \textit{UBVRI} colour filters, attached to each observation channel, enable two simultaneous colour measurements. Uncertainties in the calculated polarization are largely dependent on photon statistics, which is expected to be relatively small since \pks{} is a bright source ($m_B = 13^m.6$; \citealt{Pica1988}).

\begin{figure}
	\includegraphics[trim = 2.1cm 6.8cm 1.3cm 7.4cm, clip, width=\columnwidth]{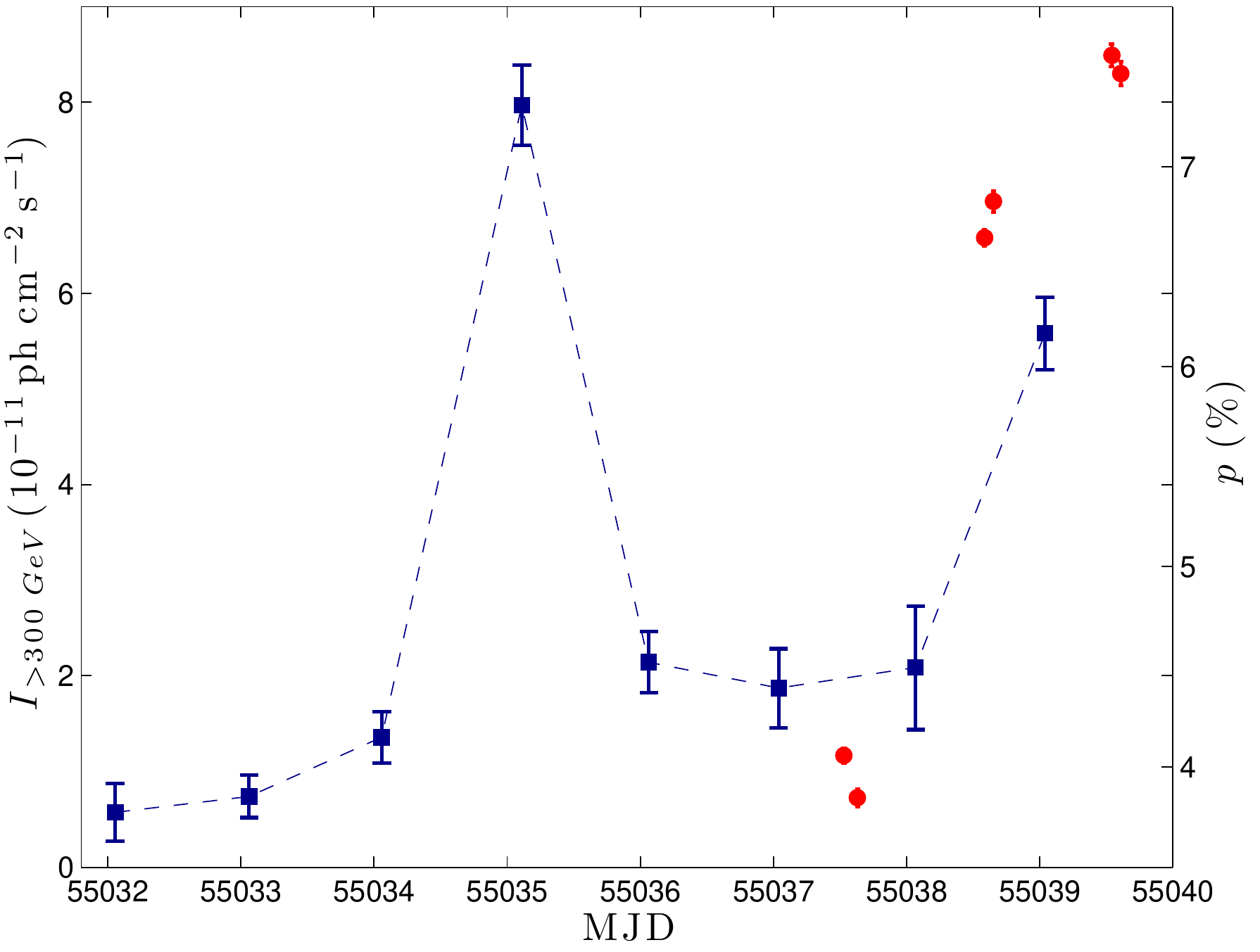}
    \caption{The \gray{} light curve of \pks{} for photons with energies $>300$ GeV in July 2009, as seen by the \hess{} \citep{HESS2014}. The \gray{} intensity \If, measured in \phcms, is represented by the filled circles. The \If--band polarization degree \p{} (\%) is superposed on the \gray{} light curve and is symbolized by the filled red squares. Two prominent increases in the \gray{}flux are detected. The first peak occurs on $\mjd=55035$, at $I=(8.0\pm 0.4)\times 10^{-11}$ \phcms. Another increase in \gray{} flux, from $(1.9\pm 0.4)\times 10^{-11}$ \phcms{} to $(5.6\pm 0.4)\times 10^{-11}$ \phcms{} is seen between $\mjd=55038$ and 55039.}
    \label{fig:obs}
\end{figure}

The intra-day variability (IDV) of the source was monitored from 25 -- 27 July 2009 for a total of 9.23 hours using a time-resolution of 5 minutes. The linear and circular polarization of the source was recorded in the \If-- and \Bf--band. A summary of the daily average of the polarization degree and the electric vector position angle (EVPA), or polarization angle\footnote{The polarization angle is measured from the North Celestial Pole towards the Celestial East.}, of the \If-- and \Bf--band is presented in Table \ref{tab:obs}. The columns list the Modified Julian date (MJD), duration ($T_\text{obs}$) and the polarization degree and polarization angle of the source. Table \ref{tab:obs} demonstrates an increase in the polarization degree from approximately 4\% to 8\% between MJD = 55037 to 55038, accompanied by a decrease in the polarization angle from 88\dgr{} to 68\dgr. Historically, the polarization degree of \pks{} typically ranges from $2\%-10\%$ on timescales of a few days \citep{Tommasi2001}, the EVPA displays a systematic decrease on timescales exceeding ten years \citep{Dominici2008}, which is consistent with our findings.    
\begin{table}
	\centering
	\caption{The mean daily optical polarization of \pks{} in July 2009.}
	\label{tab:obs}
	\begin{tabular}{cccc} % four columns, alignment for each
		\hline
		MJD & $T_\text{obs}$ (min)	&\p{} (\%) &$\theta$ (\dgr)\\
		\hline
		55037	&290	&$3.7\pm 0.3$	&$88\pm 2.5$\\
		55038	&195	&$7.0\pm 0.3$	&$67\pm 1.0$\\
		55039	&197	&$8.3\pm 0.7$	&$68\pm 0.5$\\
		\hline
	\end{tabular}
\end{table}
Figure~\ref{fig:obs} illustrates the change in the polarization degree. The \gray{} light curve as recorded by the \hess{} for photons with energies $> 300$ GeV, is superimposed. The increase in the fractional polarization is associated with an increase in \gray{} activity, with two prominent flares being detected in the \gray{} flux, \If, \citep{HESS2014}. The first flare peaks at $I=(8.0\pm 0.4)\times 10^{-11}$ \phcms{} and is located at MJD = 55035. A subsequent rise is then seen in the \gray{} flux between $\mjd=55038$ and 55039, from $I=(1.9\pm 0.4)\times 10^{-11}$ \phcms{} to $I=(5.6\pm 0.4)\times 10^{-11}$ \phcms, which could be indicative of a developing flare. However, this cannot be verified due to a lack of observations past $\mjd=55039$. Note that the second rise in \gray{} flux is accompanied by a similar rise in the polarization degree by about one day.

%\begin{figure}
%\includegraphics[trim = 2.1cm 11.cm 2.5cm 11.5cm, clip, width=\columnwidth]{polvar_I_all.pdf}
%\caption{The intra-day variability of the \If--band polarization degree, $p$ in \%, of \pks{} averaged in 5 min intervals.}\label{fig:polvar}
%\end{figure}

%\begin{figure}
%\includegraphics[trim = 2.1cm 11.cm 2.5cm 11.5cm, clip, width=\columnwidth]{polangvar_I_all.pdf}
%\caption{The intra-day variability of the position angle of the electric vector, $\theta$ in \dgr.}\label{fig:polangvar}
%\end{figure}

\section{Quasi-Periodic Oscillations}\label{sect:analysis}
The intra-day variations in the optical polarization was investigated by analyzing the high time resolution data taken over the course of each day. The number of photons collected in the \If--band were typically an order of magnitude higher compared to the \Bf--band. Since the measurement errors are dominated by photon statistics, the \Bf--band observations display larger measurement errors. More robust measurements were recorded in the \If--band, which is what we use in the following analysis. 

\begin{figure}
\includegraphics[trim = 1.9cm 6.95cm 2.25cm 7.4cm, clip, width=\columnwidth]{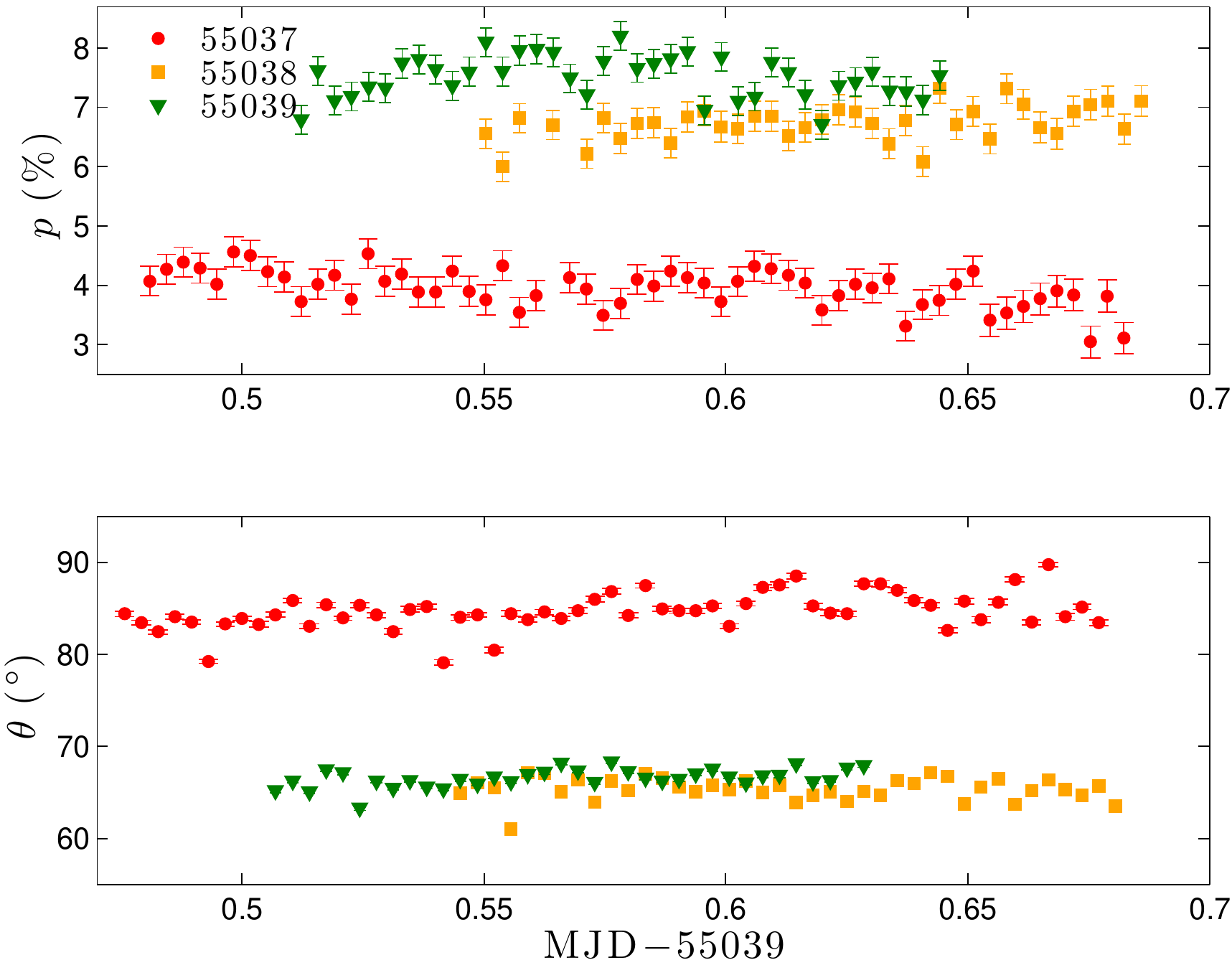}
\caption{The intra-day variability of the \If--band polarization (binned to 5 min) of \pks{} on $\mjd=55037$ (circles), 55038 (square markers) and 55039 (triangles). The top panel displays the polarization degree $p$ (\%), while the position angle of the electric vector $\theta$ (\dgr) is shown in the bottom panel. The polarization degree appears to modulated by a periodic component on $\mjd=55037$.}\label{fig:polvar}
\end{figure}

Inspection of the IDV on $\mjd = 55037$ over the 290 minutes of continuous observation (see Fig.~\ref{fig:polvar}) suggests that the amplitude of the polarization degree is modulated on a timescale of $\sim 30$ min. These short-term amplitude modulations are superposed on a slowly decreasing baseline. The oscillations thus precede the onset of the increase in the optical polarization degree seen in Fig.~\ref{fig:obs}. It is noteworthy such short-term modulations are not seen on $\mjd = 55038$ and 55039, after the onset of the rise in the polarized intensity and the second \gray{} flare.
%\textbf{Caveat: Visually, there appears to be some indication of epi-cyclic amplitude modulation in the polarization degree on the second night. However, due to limited time span/ duration of data on second and third night we are cannot reliably perform statistical analysis. Although the periodogram and PSD show presence of peak, detected at 95\% significance in \Q{} on second night, with T=12.5.}

\begin{figure}
\includegraphics[trim = 1.9cm 9.6cm 2.25cm 9.9cm, clip, width=\columnwidth]{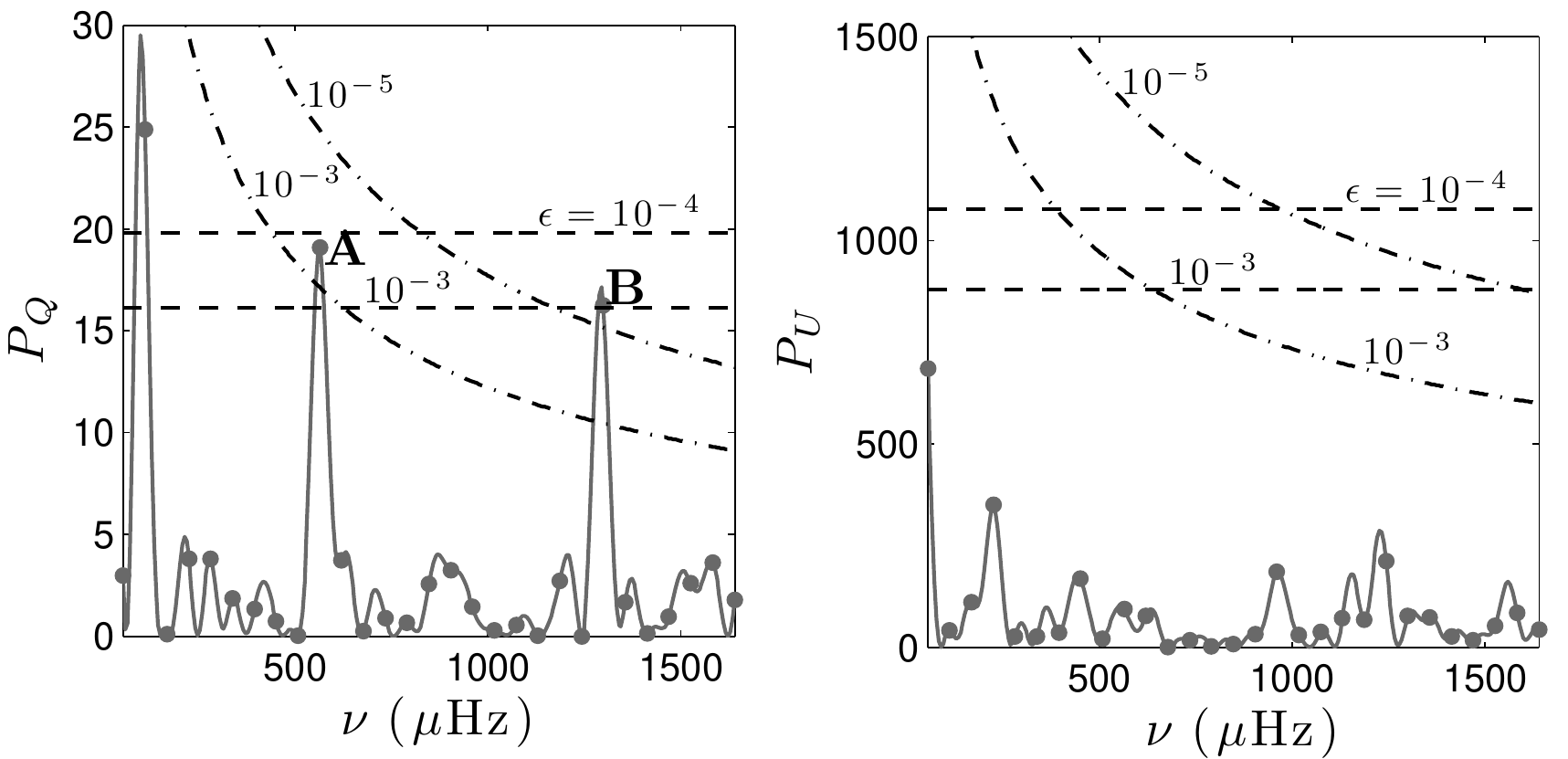}
\caption{The periodogram of the polarized flux, in units of $(\text{rms}/\text{mean})^2$ per Hz. The periodogram of the Stokes \Q{} flux, $P_Q$, is presented in the left panel, while the periodogram of the Stokes \U{} flux, $P_U$, is shown on the right. The filled circles represent the periodogram ordinates at the Fourier frequencies. Two prominent peaks are detected at $\nu_A=565\pm 7$ \mHz{} and $\nu_B=1293\pm 6$ \mHz, which are labelled as \textbf{A} and \textbf{B}, respectively. The false alarm probabilities for white noise, indicated by the dashed lines, and red noise, indicated by the dotted-dashed lines, are superposed.}\label{fig:powspec}
\end{figure}

To test for the presence of periodic components in the optical polarization we computed the periodogram, which measures the amount of variability power as a function of temporal frequency, normalized to the mean of the light curve, in units of $(\text{rms}/\text{mean})^2$ per Hz \citep{Schuster1898, Vaughan2005}. 
%The normalisation is chosen such that the units of the peri- odogram are (rms/mean)2 Hz−1 (where rms/mean is dimension- less) and summing the periodogram over positive frequencies gives the sample variance in fractional units \citep{Vaughan2005}. 
The periodogram was oversampled by a factor of 10 relative to the grid defined by the Fourier frequencies $\nu_k=k/N\Delta T$ to provide good sampling of the spectrum, where $N$ is the number of observations and $\Delta T$ is the sampling interval. The periodogram of the Stokes \Q{} and \U{} polarized intensities is displayed in Fig.~\ref{fig:powspec}. Two prominent peaks are detected at $565\pm 7$ \mHz{} and $1293\pm 6$ \mHz, labelled as A and B, respectively. Each peak signal is well-described by a Gaussian function, for which the uncertainty in the position of the peak is given by $\sigma_{\nu_0}= \text{FWHM}\bigr/\left(2\sqrt{2\ln 2}\times \sqrt{snr}\right)$ \citep{Condon1997}, where FWHM is the full-width half-maximum of the peak and $snr$ is its signal-to-noise ratio. The period corresponding to peak A is $T=29.4$ min and $T=12.9$ for peak B. No dominant peaks are detected in the Stokes \U{} periodogram. The lack of detection in \U{} is consistent with a variation in polarized intensity (as opposed to angle), as the observed polarization position angle $\theta\approx 90$\dgr{} (see Table~\ref{tab:obs} and Fig.~\ref{fig:polvar}), places virtually all the polarized emission in \Q.
%The model spectrum in each case was determined by generating $10^4$ light curves from a white noise distribution with the same mean, variance and sampling rate as the observations. Each periodogram was fitted with a white noise model. The resulting distribution in variance allowed a more accurate determination of the noise variance.

\begin{figure}
\includegraphics[trim = 2.7cm 10.4cm 2.35cm 10.9cm, clip, width=\columnwidth]{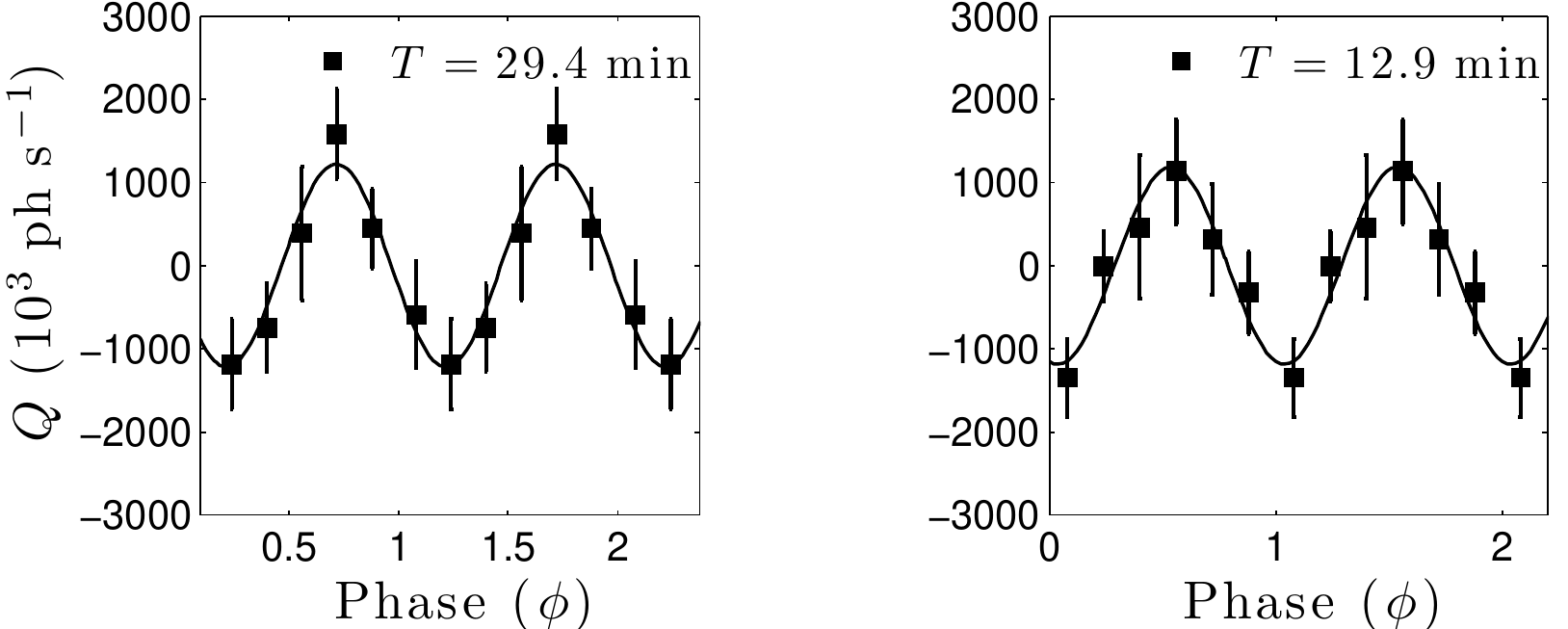}
\caption{The Stokes \Q{} light curve folded to $T=29.4$ min and $T=12.9$ min, averaged into six phase bins. Two cycles are plotted for clarity. A cyclic trend appears to be present for both trial periods. The solid line represents the best-fit sinusoid for each case.}\label{fig:phasefold}
\end{figure}

Figure~\ref{fig:phasefold} displays the Stokes \Q{} light curve folded to the $T=29.4$ min, the period of spectral component \textbf{A} and $T=12.9$ min, the period of spectral component \textbf{B}. Two cycles are plotted for clarity. Although the amplitude modulation is small, a cyclic trend appears to be present for both periods. A non-linear least-squares fit to the data yields a best-fit sinusoid that agrees well with the phase-folded diagram, with a period of $T=29.5$ min for signal A and $T=12.9$ min for signal B.

\begin{figure}
\includegraphics[trim = 1.5cm 6.8cm 2.15cm 7.3cm, clip, width=\columnwidth]{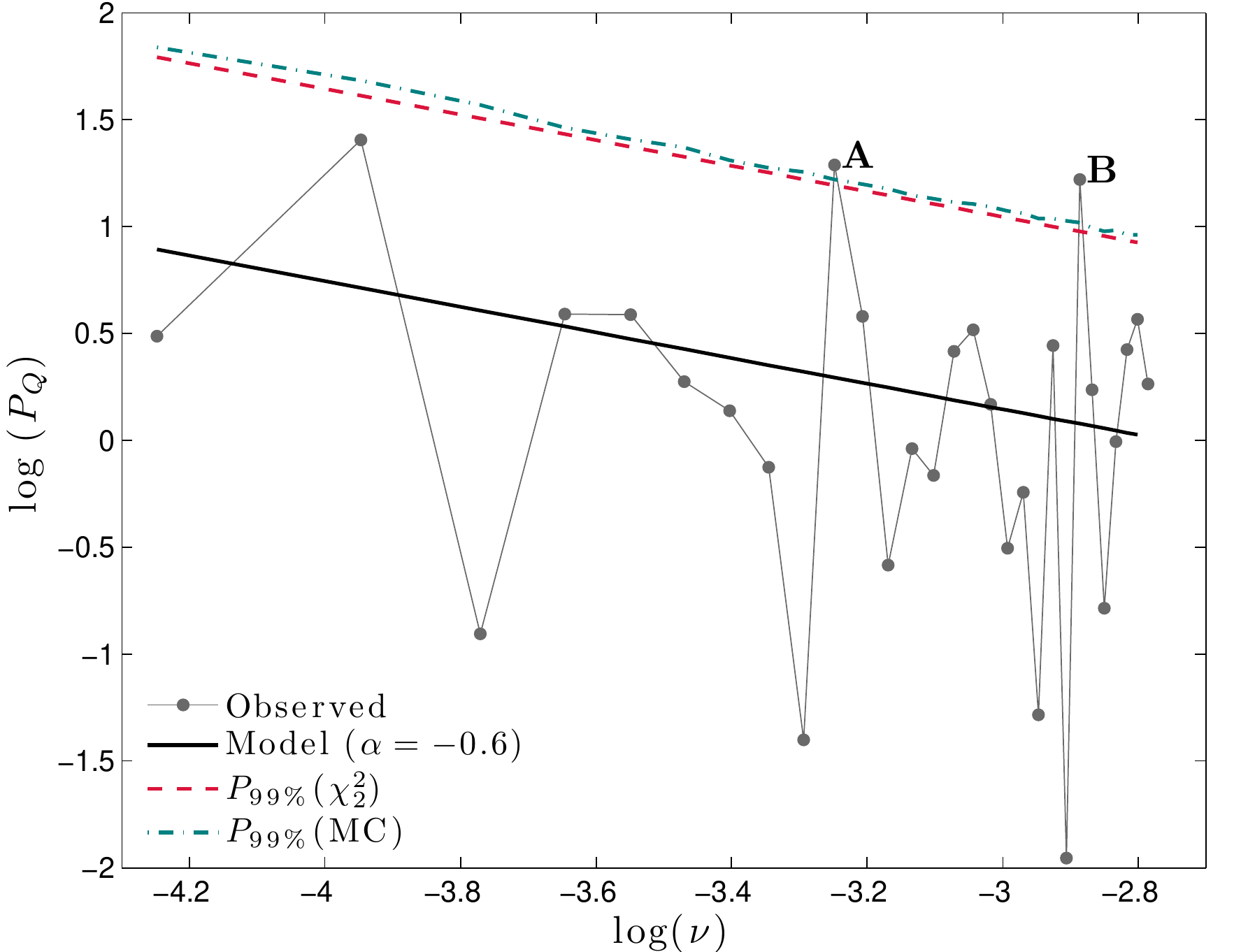}
\caption{The power spectrum of the Stokes \Q{} intensity, normalized to $(\text{rms}/\text{mean})^2$ per Hz. The solid line represents the best-fit power-law with a slope $\alpha=0.59\pm 0.3$. The 99\% confidence limit of the noise is represented by the dashed line. The robustness of this result was tested by generating $10^4$ random light curves with the same mean, variance and sampling rate as the observations from an underlying continuum with $\alpha=0.6$. The 99\% confidence limit of the simulated power distribution is indicated by the teal dotted-dashed line.}\label{fig:PSDfit}
\end{figure}

The statistical significance of the periodogram was determined by comparing the observed peak power to the probabilities of deflections arising from a noise continuum $S=N\nu^{-\alpha}$ (or red noise), where $\alpha$ is the spectral index, or slope, and $N$ is the flux normalization of the spectrum. The model spectrum is obtained by fitting a power-law to the observed periodogram using the least-squares (LS) method described in \citealt{Vaughan2005}. The result of the fit is displayed in Fig.~\ref{fig:PSDfit}, which shows the Stokes \Q{} power spectrum on a logarithmic axes. The slope is consistent with white noise ($\alpha = 0$), but a formal fit to the spectrum yields a de-biased best-fit line with $\alpha=0.59\pm 0.3$ (represented by the solid line). If $S$ represents the true power spectral density, then the ratio $\gamma = 2P/S$ is scattered as a \chisq -- distribution with two degrees of freedom and we can define confidence limits for the noise. The upper 99\% confidence limit of the underlying continuum was calculated. Fig.~\ref{fig:PSDfit} shows that the observed peaks are detected above the 99\% confidence limit (represented by the red dashed line), with the chance probability of detecting peak A $\text{Pr}[\gamma>\gamma_\epsilon]=3.1\times 10^{-4}$ (corresponding to $\sim 3\sigma$) and $3.4\times 10^{-6}$ for peak B, given the number of frequencies that were examined. The false alarm probabilities for the power-law model, and for white noise, are superposed on the periodogram in Fig.~\ref{fig:powspec}. The figure illustrates that the observed peaks exceed the $\epsilon=10^{-3}$ false alarm threshold of white noise. %Fig.~\ref{fig:powspec} shows that peak A and peak B are statistically significant, with both peaks exceeding the $10^{-3}$ false alarm threshold of white noise. For red noise, peak A is detected above $\epsilon=10^{-3}$, while peak B exceeds the $10^{-4}$ false alarm threshold. No significant peaks are detected in \U.%Oversampling the periodogram and assuming n trials will tend to overestimate the significance of peaks in the periodogram.

The robustness of this result was verified by generating $10^4$ random light curves with the same mean, variance and sampling rate as the observations, following the procedure described in \citealt{TimmerKonig1995}. The resulting distribution of power for an underlying continuum with $\alpha=0.6$ (the best-fit slope to the data) was recorded and the 99\% and 99.7\% confidence levels computed. The 99\% confidence level from the simulated distribution is indicated by the teal dotted-dashed line in Fig.~\ref{fig:PSDfit}. Although the simulation shows that the LS test overestimates the significance of the periodic signal, peak A and peak B remain statistically significant, with peak A detected at $>99$\% significance, while peak B is detected at $>99.7$\% significance. 
% , where the probability of obtaining a higher value by chance the observed power, or false alarm probability

To conclude, the periodogram of the polarized flux is consistent with the presence of a periodic component at $T\sim 13$ min, with $>99.7$\% significance, and $T\sim 30$ min, which was found to be nominally significant at $>99$\%.
%I performed the same test for white noise and for noise with slope equal to 1. For the former, only peak A is statistically significant, with the peak power exceeding the 99\% confidence level. For the latter, both peaks are statistically significant above the 99.7\% confidence limit. 

\section{Discussion}\label{sect:discussion}
The optical emission in blazars is dominated by synchrotron radiation from relativistic electrons being accelerated in the jet's magnetic field. The polarization is determined by the degree of order and direction of the magnetic field, and the viewing angle of the observer. Since radiation from the jet is highly Doppler boosted, even for small changes in the viewing angle, significant brightness variations will be observed. Fast variations, such as those observed in the optical polarization of \pks, therefore most likely originate in the Doppler boosted jet, from an emission region with a size which is comparable to the gravitational radius of the central engine. Using the timescale of the fastest periodic component as the variability timescale $t_\text{var}\approx 13$ min, a redshift $z=0.116$ \citep{Falomo1993} and a Doppler factor $\delta\approx 30$ \citep{Katarzynski2008}, yields an upper limit to the size of the emission region $R \lae c \delta t_\text{var}/(1+z) \approx 10^{15}$ cm.

Quasi-periodic variations can arise if an emission feature with physical properties distinct from the ambient jet medium, such as a relativistic shock, propagates down a jet which contains quasi-helical structures, either in the magnetic field or electron density (e.g., \citealt{Qian1991, Romero1995}).  A relativistic shock propagating down such an inhomogeneous jet will result in successive peaks in the amplitude of the total and polarized emission at the viewing angle that provides the maximum boosting for the observer (e.g., \citealt{CamenzindKrockenberger1992, GopalWiita1992}).

Turbulence behind the shock can also explain the presence of short-lived QPOs \citep{Marscher1992}. For such turbulent flows, the largest eddies will dominate the variability and the period of the period will be determined by their turnover times. The presence of one or more such turbulent cells behind the same shock also provides a possible explanation for the presence of multiple QPOs in the same signal.% (e.g., \citealt{Lachowicz2009,Rani2009,Rani2010}).} 
%Observational evidence for the existence of edge-brightened and non-axisymmetric structures have been seen in the resolved radio images of some AGN, e.g., M87, Ly et al. 2007; Mkn 501, Piner et al. 2009.
%A relativistic shock propagating down such a perturbed jet will induce significantly increased emission at the locations where the shock intersects a region of enhanced magnetic field and/or electron density corresponding to such a nonaxisymmetric structure. Thanks to the extreme sensitivity of Doppler boosting to viewing angle, very substantial changes in the amplitude (and polarization) of radio and optical jet emission will be seen by an observer at fixed angle to the jet axis as the most strongly emitting region effectively swings past the observer (e.g., \citealt{CamenzindKrockenberger1992, GopalWiita1992}). 
  
%Quai-peiodic changes in the viewing angle vs inhomogeneities in jet argument- romero 1995
%Instabilities in jets just might be able to excite such helical modes capable of yielding fluctuations that are ob- served to occur on the time-scale seen in PKS 2155−304 (e.g., Romero 1995). Or they could arise as the jet plasma is launched in the vicinity of SMBH and thus actually originate in the accre- tion flow but become amplified in the jet.

\section{Conclusions}\label{sect:conclusions}
Monitoring of the intra-day optical polarization of \pks{} revealed the first evidence for quasi-periodic oscillations in the polarized flux of an AGN. The periodogram showed the existence of a periodic component at $T\sim 13$ min, detected at $>99.7$\% significance, and $T\sim 30$ min, which was found to be nominally significant at $>99$\%. The period of these oscillations are similar to the $T\approx 15$ min QPO seen in the optical light curve of the blazar S5 0716+714 \citep{Rani2010}, which provides further support for the existence of such short timescale periodicities in these sources. 

Gamma-ray observations of the source over the same period showed that it was in a high state of activity, experiencing two \gray{} flares at very high energies (VHE, photons with energies exceeding a few GeV). The first simultaneous optical polarization measurements of the source during a high-state was recorded for the latter flare. Although the physical cause of this QPO is unclear, comparison with the VHE light curve showed that \pks{} experienced a \gray{} flare two days before the appearance of the QPOs and another peak two days later. The oscillations could therefore be related either to the late phase of post flare activity of the first flare, which is supported by the decreasing trend observed over the night (as opposed to increasing trend seen on the following nights, see Fig.~\ref{fig:polvar}) or due to transition between two different gamma-ray flares. Since blazar emission is dominated by synchrotron emission at low energies, which also leads to polarization, and inverse Compton emission at high energies, which both arise from the relativistic jet, this could suggest that the observed QPOs are part of a longer-lived phenomenon within the jet of \pks.
%the emission is dominated by synchrotron emission and inverse Compton emission which both arise from relativistic jets. The quasi periodic variations are mostly caused by blobs propagating in a helical jet. The effect is similar to that of a jet whose direction changes. When the viewing angle is small, the source brighter, and vice versa. In this case, one would expect to see a rather exact variability pattern for some periods until the blob vanishes, and then a repetition with a new blob. \citep{Zhang2014}

\section*{Acknowledgements}
%The Acknowledgements section is not numbered. Here you can thank helpful colleagues, acknowledge funding agencies, telescopes and facilities used etc. Try to keep it short.
%The authors would like to acknowledge the support of their host institutions. 
The authors would like to thank the anonymous referee for many suggestions that significantly improved this paper. 

We gratefully acknowledge the support of the South African National Research Foundation (NRF) and the Department of Science and Technology (DST) through the South African Square Kilometre Array Office. N.W.P. also thanks the Science Faculty at the University of Cape Town (UCT) and Centre for Space Physics at North-West University (NWU) for their support. The authors also thanks the \hess{} Collaboration for taking the \gray{} observations, as well as Markus B\"{o}ttcher for stimulating discussions. 

%Not sure if we need an acknowledgement to the HESS data base which you used for the Gamma Ray points.

%%%%%%%%%%%%%%%%%%%%%%%%%%%%%%%%%%%%%%%%%%%%%%%%%%



%\bibliographystyle{mnras}


%%%%%%%%%%%%%%%%%%%%%%%%%%%%%%%%%%%%%%%%%%%%%%%%%%

%%%%%%%%%%%%%%%%% APPENDICES %%%%%%%%%%%%%%%%%%%%%

%\appendix


%If you want to present additional material which would interrupt the flow of the main paper, it can be placed in an Appendix which appears after the list of references.

%%%%%%%%%%%%%%%%%%%%%%%%%%%%%%%%%%%%%%%%%%%%%%%%%%

% Don't change these lines
\bsp	% typesetting comment
\label{lastpage}
\end{document}

%% file: newcommands.tex
\newcommand{\gae}{\lower 2pt \hbox{$\, \buildrel {\scriptstyle >}\over {\scriptstyle
\sim}\,$}}
\newcommand{\lae}{\lower 2pt \hbox{$\, \buildrel {\scriptstyle <}\over {\scriptstyle
\sim}\,$}}

\newcommand{\mjd}{\text{MJD}}

\newcommand{\pks}{PKS 2155--304}

\newcommand{\mHz}{$\mu$Hz}
\newcommand{\msun}{M$_\odot$}

\newcommand{\dgr}{$^\circ$}
\newcommand{\Q}{\textit{Q}}
\newcommand{\U}{\textit{U}}
\newcommand{\chisq}{$\chi^2_2$}

\newcommand{\Bf}{\textit{B}}

\newcommand{\If}{\textit{I}}

\newcommand{\hess}{H.E.S.S.}

\newcommand{\p}{\textit{p}}

\newcommand{\phcms}{ph~cm$^{-2}$~s$^{-1}$}

\newcommand{\xray}[1]{$X-$ray #1}
\newcommand{\gray}[1]{$\gamma-$ray #1}

\newcommand{\til}{$\sim$}